\documentclass[a4paper,nobibnotes,nofootinbib]{revtex4}

\usepackage{amsmath, epsfig}

\newcommand{\be}{\begin{equation}}
\newcommand{\ee}{\end{equation}}
\newcommand{\bea}{\begin{eqnarray}}
\newcommand{\eea}{\end{eqnarray}}

\newcommand{\f}{\frac}

\newcommand\lr[1]{{\left({#1}\right)}}

\begin{document}

\title{Jet quenching in the strongly-interacting quark-gluon plasma}
\author{C. Marquet}\email{cyrille.marquet@cea.fr}
\affiliation{Institut de Physique Th{\'e}orique, CEA/Saclay, 91191 Gif-sur-Yvette cedex, France}
\author{T. Renk}\email{thorsten.i.renk@jyu.fi}
\affiliation{Department of Physics, P.O. Box 35, 40014 University of Jyv\"{a}skyl\"{a}, Finland}
\affiliation{Helsinki Institute of Physics, P.O. Box 64, 00014 University of Helsinki, Finland}

\begin{abstract}

We propose a hybrid model for medium-induced parton energy loss, in which the hard scales in the process are treated perturbatively, while the soft scales which involve strong coupling dynamics are modeled by AdS/CFT calculations. After fitting a single parameter on $R_{AA}$ for central Au+Au collisions, we are able to predict different observables like $R_{AA}$ and $I_{AA}$ as a function of centrality and reaction plane. We obtain a consistent picture of how jet quenching is modified if the quark-gluon plasma is strongly interacting, and we provide quantitative predictions.

\end{abstract}

\maketitle

\section{Introduction}

Understanding the quark-gluon plasma (QGP) created in nucleus-nucleus collisions
\cite{brahms,phobos,star,phenix} at the Relativistic Heavy Ion Collider (RHIC) is a challenging task. Many measurements have been carried out to provide insight on the properties of that dense QCD matter, yet a few years after the discovery, the question whether the QGP formed at RHIC
is weakly or strongly coupled remains unanswered. On the one hand, some features of the plasma
are naturally understood from strong cross-sections, and even fit quantitatively within a
perfect-liquid picture, like the large elliptic flow
\cite{Dusling:2007gi,Song:2007ux,Luzum:2008cw}. On the other hand, one would like to establish a global picture and be able to also answer the question using hard probes. While qualitatively the medium is opaque to energetic particles, implying strong interactions, a quantitative strong-coupling description of jet quenching is missing.

Hard probes are believed to be ideal processes to study the properties of the QGP, they are thought to be understood well enough to provide clean measurements. Observables built to measure medium effects on particle production, like the nuclear modification factors $R_{AA},$ are not always easily reproduced in perturbative QCD (pQCD) calculations, even for hard production. For instance in older computations, to reproduce light-hadron $R_{AA}$'s within the pQCD framework of medium-induced energy loss \cite{Baier:1998kq,Wiedemann:2000za,Gyulassy:2000er}, the value of the so-called jet quenching parameter $\hat{q}$ is adjusted to $5-10\ \mbox{GeV}^2/\mbox{fm}$ \cite{Eskola:2004cr}. It is otherwise estimated to be smaller ($1-3\ \mbox{GeV}^2/\mbox{fm}$) for a weakly-coupled plasma at RHIC temperatures. A more recent comparison study of different pQCD approaches within the same medium model \cite{Bass:2008rv} shows that assumptions about how
$\hat{q}$ depends on thermodynamical parameters play a crucial role and that in principle pQCD based models may also work with a low $\hat{q}$. While our current understanding of the pQCD picture must be still be improved
, it is unclear if the pQCD approach can describe the suppression of high$-P_T$ particles at RHIC.

The goal of this letter is to quantify the expected modifications of the fully-perturbative radiative energy loss results in QCD, if the plasma is strongly coupled. Addressing the strong coupling dynamics in QCD is an outstanding problem, lattice simulations remain the only method
to obtain quantitative results, but are inefficient when real-time dynamics is required, which is the case when analyzing jet quenching. However for a class of non-abelian thermal gauge theories,
the AdS/CFT correspondence \cite{Maldacena:1997re,Witten:1998qj,Gubser:1998bc} provides an alternative: it allows to investigate the strong coupling regime in the large-$N_c$ limit, essentially by mapping the quantum dynamics of the gauge theory into analytically-tractable classical gravity dynamics in the fifth dimension of a curved anti-deSitter (AdS) spacetime. 

We propose a hybrid model for medium-induced parton energy loss, in which the hard scales in the process are treated perturbatively, as in the standard pQCD radiative energy loss calculations,
while the interaction with the plasma which involves strong-coupling dynamics is modeled by AdS/CFT calculations, for the $\mathcal{N}\!=\!4$ super-Yang-Mills (SYM) theory. While this theory is quite different from QCD (it is highly supersymmetric and conformal), it has become a popular approach to assume that the SYM plasma approximates well the QCD plasma, just above the critical temperature $T_c,$ where the conformal anomaly in QCD is small. Information about real-time dynamics within a thermal background can be obtained with this setup. The best-known example is the calculation of the shear viscosity to entropy ratio \cite{Policastro:2001yc,Kovtun:2004de}, which in the strong coupling limit holds for any gauge theory with a gravity dual.

After a bare parton is created in a hard process, it starts to build its wave function with quantum fluctuations. The presence of the medium prevents the parton to become fully dressed as it would in the vacuum: long-range fluctuations, typically with virtuality less than the plasma temperature
$T,$ are screened out of the wave function. The parent parton then loses the energy that those fluctuations carry into the medium. We would like to compute the properties of this medium effect in QCD. In the high parton energy limit, the energy is transfered into the medium from the perturbative part of the parent parton wave function. Then we explicitly know what the quantum fluctuations are, quarks and gluons, and their dynamics can be computed from first principles
\cite{Kovner:2003zj}. On the contrary, if one were to probe the soft part of the wave function, then we wouldn't know how to deal with the fluctuations and their dynamics in QCD.

Having in mind the description of high$-P_T$ hadrons in heavy-ion collisions, which come from the fragmentation of energetic partons, the authors in \cite{LRW} proposed to use the
Baier-Dokshitzer-Mueller-Peigne-Schiff-Zakharov (BDMPS-Z) energy loss formalism
\cite{BDMPS,Zakharov} to calculate the gluon emission process, which involves $\alpha_s(p_T^2),$ while describing the interaction of the emitted gluons with the strongly-coupled medium with
AdS/CFT calculations. Their motivation is that, in the high-parton energy limit, the BDMPS-Z analysis of medium induced parton energy loss is under control, and whether the plasma is weakly-coupled or not gluon radiation is the dominant mechanism \cite{Mueller:2008zt}. Moreover, applying strong-coupling techniques to the entire radiation process would be inappropriate, because QCD is asymptotically free, and the calculation of the gluon distribution in the parent parton wave function is correctly treated with perturbative methods.

In the so-called multiple soft scattering approximation of the BDMPS-Z calculation, the properties of the energy loss are expressed in terms of the transport coefficient $\hat{q},$ a constant rate at which gluons in the parent parton wave function pick up transverse momentum squared
$dp_\perp^2/dt=\hat{q}\propto T^3.$ Instead of evaluating $\hat{q}$ perturbatively, the authors of \cite{LRW} proposed to evaluate this coefficient at strong coupling. Unfortunately, at strong coupling, the energy loss cannot be expressed in terms of a constant $dp_\perp^2/dt$
\cite{Dominguez:2008vd,Marquet:2008kr}. However, an interpretation based on the multiple soft scattering approximation in the BDMPS-Z calculation is still possible, but instead with multiple scatterings characterized by the constant rate $d|p_\perp|/dt\propto T^2.$ Our strategy is to express the properties of the energy loss in terms of the accumulated $p_\perp^2$ as in the BDMPS-Z calculation, and then to evaluate this accumulated transverse momentum squared, that we shall denote $Q_s^2,$ at strong coupling: $Q_s^2=T^4L^2$ (instead of $Q_s^2=\hat{q}L$ at weak coupling), where $L$ is the extent of the medium. Our hybrid model is then completed by a realistic account of the plasma geometry and expansion, as in \cite{Renk:2006pk}.

The plan of the letter is as follows. In Section II, we recall the different pieces of the BDMPS-Z medium-induced radiative energy loss calculation, as well as the physical picture emerging from such a weak-coupling analysis. We also explain what part of the calculation should be modified in the case of a strongly-coupled plasma, and propose a substitution inspired by AdS/CFT results. In Section III, we discuss how the plasma geometry and expansion are implemented in order to obtain meaningful predictions. In Section IV, we present results for $R_{AA}$ and $I_{AA}$ at high $P_T$ as a function of the reaction plane, observables for which the stronger path-length dependence of the energy loss at strong coupling, compared to weak coupling, produces sizable differences. Section V is devoted to conclusions and outlook.

\section{Medium-induced parton energy loss}

We denote the energy of the hard parton $E,$ and work in the large parton energy limit. We also denote $\omega$ and $k_\perp,$ the energy and transverse momentum of the virtual gluons in the parent parton wave function (note that we are using $p_T$ to denote transverse momenta with respect to the collision axis, and $p_\perp$ with respect to the jet axis). What prevents the radiated gluons to have an arbitrarily large energy, is that their coherence time $t_c$ should be smaller than the length of medium $L$ the parent parton goes through before exiting the medium:
\be
t_c=\omega/k_\perp^2\ , \quad \omega < L\ k_\perp^2\ .
\ee
The transverse momentum squared, or virtuality, of the radiated gluons is also limited: the interaction with the plasma cannot overcome an arbitrary large virtuality and put on shell any fluctuations. Let us denote $p_\perp^2(t_c)$ the transverse momentum squared acquired by a gluon of coherence time $t_c,$ then only the gluons with virtuality $k_\perp^2<p_\perp^2(t_c)$ can be radiated. For those emitted gluons which dominate the energy loss of the parent parton, we shall denote their virtuality and energy
\be
Q_s^2\equiv p_\perp^2(L)\ ,\quad\mbox{and}\quad \omega_m=L\ Q_s^2\ .
\ee
The saturation momentum $Q_s$ depends on the temperature $T$ and length $L$ of the plasma, and the precise way it depends on it is different if the plasma is weakly coupled or strongly coupled.

The first part of the BDMPS calculation is to determine the energy distribution of the radiated gluons $\omega dI/d\omega,$ from which the total energy loss $\Delta E,$ as well as the quenching weight
\be
\label{QuenchingWeight}
P(\Delta E)=\sum_{n=0}^\infty\f1{n!}
\left[\prod_{i=1}^n\int d\omega_i\ \f{dI(\omega_i)}{d\omega}\right]
\delta\lr{\Delta E -\sum_{i=1}^n\omega_i}\exp\lr{-\int d\omega\ \frac{dI}{d\omega}}
\ee
necessary to compute $R_{AA},$ can be computed \cite{Salgado:2003gb}. The saturation momentum is what determines the properties of the energy loss, as the quenching weight are expressed in terms of $Q_s^2$ and $\omega_m.$ The other part of the BDMPS calculation assumes that the plasma is made of thermalized weakly-interacting quarks and gluons, which allows to determine $Q_s^2$ and
$\omega_m$ as functions of the plasma temperature and length.

\subsection{In a weakly-coupled plasma}

At weak-coupling, medium-induced gluon radiation is due to multiple scatterings of the virtual gluons \cite{BDMPS}: if, while undergoing multiple scattering, the virtual gluons pick up enough transverse momentum to be put on shell, they become emitted radiation. The accumulated transverse momentum squared picked up by a gluon of coherence time $t_c$ is
\be
p_\perp^2(t_c)=\mu^2 \f{t_c}{\lambda}\equiv\hat{q}\ t_c\ ,
\quad\mbox{which gives}\quad Q_s^2=\hat{q}L\ ,
\ee
where $\mu^2$ is the transverse momentum squared picked up in each scattering, and $\lambda$ is the mean free path. These medium properties are involved through the ratio $\hat{q}=\mu^2/\lambda,$ this is the only relevant information about the medium. In terms of the temperature $T,$ one has
$\hat{q}\sim\alpha_s T^3.$ The total energy lost by the quark is therefore
$\Delta E\propto\alpha_s\hat{q}L^2,$ where the factor $\alpha_s$ comes from the probability for the gluonic fluctuation in the wave function, and essentially counts the number of gluons which lose the energy $\hat{q} L^2.$

An important quantity which enters the calculation of $\omega dI/d\omega$ is the scattering amplitude of a color singlet dipole off the plasma, which at weak coupling is given by
\be
N(r)=1-e^{-\hat{q} L r^2}\ .
\ee
This explains why we denoted the accumulated $p_\perp^2$ by $Q^2_s:$ this sets the scale at which the scattering becomes strong. The dipole is made of the parent parton and its antiparticle (it could be a $q\bar{q}$ dipole or a $gg$ dipole, the difference is a color factor absorbed in
$\hat{q}$). In the so-called multiple soft scattering approximation that we are considering here, the scattering amplitude for small dipoles in a single scattering
$N(r)\sim\hat{q}\lambda r^2=\mu^2r^2$ is explicitly used. This allows to obtain the quenching weight as a function of $Q_s^2$ and $\omega_m$ \cite{Wiedemann:2000tf}. In phenomenological calculations, the geometry and expansion of the plasma are taken into account by integrating over the parton trajectory $\xi$ in the medium \cite{Renk:2006pk}:
\be
Q_s^2(\textbf{r}_0,\phi)=K\int_0^\infty d\xi\ T^3(\xi)\ ,\quad
\omega_m(\textbf{r}_0,\phi)=K\int_0^\infty d\xi\ \xi\ T^3(\xi)\ .
\label{weakLI}\ee
$\textbf{r}_0$ and $\phi$ denote the position of the hard process that created the parton and its direction with respect to the reaction plane. The normalization $K$ is the only parameter in the problem.

\subsection{In a strongly-coupled plasma}

We would like to address the case of a strongly-interacting quark-gluon plasma. 
We propose to model the strong coupling dynamics by replacing in the BDMPS calculation, the values of $Q_s^2$ and $\omega_m$ by their strong-coupling expressions. In other words, we assume that what is radiated into the medium still comes from the perturbative part of the energetic parton wave function, but we assume that the interaction of those virtual gluons with the medium, and the way they are freed, is governed by strong-coupling dynamics. At strong coupling the value of $Q_s$ can be estimated by a classical gravity calculation, in 5 dimensions. Indeed, using the AdS/CFT correspondence \cite{Maldacena:1997re,Witten:1998qj,Gubser:1998bc} which maps the quantum dynamics of the SYM field theory onto classical dynamics in a fifth dimension, one can evaluate the saturation momentum for this theory. We shall then use it as an input for our model. In those classical calculations, the plasma acts as a force of magnitude $T^2$ on the energetic parton and its wave function, to which correspond a string in the fifth dimension.

The work done by the plasma on a fluctuation of coherence time $t_c$ is $T^2 t_c.$
To put a quantum fluctuation on shell, this work should overcome its virtuality, therefore we identify
\be
p_\perp^2(t_c)=T^4 t_c^2\ ,\quad\mbox{which gives}\quad Q_s^2=T^4L^2
\ee
for those fluctuations which dominate the energy loss.
The expression $p^2_\perp(t_c)$ was obtained in \cite{Hatta:2007cs} with an $R-$current deep inelastic scattering (DIS) calculation. How this result translated into the length dependence of the saturation scale and the energy loss was discussed in \cite{Dominguez:2008vd,Marquet:2008kr} for an energetic heavy quark, and the main result was that at strong coupling $\Delta E\propto L^3$ instead of the $L^2$ law at weak coupling. Calculations of DIS off a finite-length strongly-coupled SYM plasma confirmed the result \cite{Mueller:2008bt,Levin:2009vj,Albacete:2008ze}. It is obtained that $Q^2_s=T^3 L/x,$ where $x$ is the Bjorken variable. In DIS off a weakly-coupled QCD plasma, the result $Q^2_s=T^3 L$ is independent of $x$, i.e. independent of the collision energy
$\sqrt{s}\sim 1/x.$ This comes from the fact that at weak-coupling, the quasiparticle which determines high-energy scattering is a spin 1 object (usually called Pomeron), and therefore
$s^{J-1}\sim 1$ (there will be an $x$ dependence via the plasma gluon distribution $xg(x)$ if one takes into account quantum evolution \cite{CasalderreySolana:2007sw}). By contrast, at strong coupling the quasiparticle which determines high-energy scattering is a spin 2 object (therefore called graviton), and $s^{J-1}\sim 1/x,$ which explains the differences between the weak- and strong-coupling saturation momenta. In the energy loss calculation, the extra $1/x$ factor is simply replaced by $t_cT,$ and for a plasma of small enough extent such that it is indeed the finite length of the plasma which limits the coherence time of the freed fluctuations, one recovers $Q_s=T^2 L.$ Finally, the $\Delta E\propto L^3$ law at strong coupling can also be obtained for gluons and light quarks, from the dynamics of falling strings \cite{Gubser:2008as,Chesler:2008uy}.

In order to make the connection with results in the literature, let us briefly consider the case of an infinite-extent plasma. Then $t_c=\omega/k^2_\perp$ is not restricted anymore, and the transverse momentum bounds become $k_\perp^4< \hat{q}\omega$ in the weak coupling case, and
$k^3_\perp<T^2 \omega$ in the strong coupling case. These, coupled with the wave function suppression for $\omega> E k_\perp/ M$ (the dead cone effect) \cite{Dokshitzer:2001zm} in the case of a heavy quark parent parton give $Q^2_s = (\hat{q}E/M)^{2/3}$ at weak coupling and
$Q^2_s = T^2 E/M$ at strong coupling for the virtualities of the most energetic freed fluctuations, whose energies are $\omega_m=EQ_s/M.$ If the plasma has a finite length, but bigger than the corresponding coherence times $E/(MQ_s),$ then the matter is effectively of infinite extent. In the strong coupling case, we recover that the saturation scale $Q_s = T\sqrt{E/M}$ in the gauge theory corresponds on the gravity side to the string world sheet horizon of the trailing string calculation \cite{Herzog:2006gh,Gubser:2006bz}, which determines the heavy quark energy loss
\cite{Chernicoff:2008sa,Beuf:2008ep}.

When computing $dI/d\omega,$ we are still using the multiple soft scattering approximation, but with the strong coupling expression for $Q_s^2.$ At the end, we substitute into the analysis of
\cite{Renk:2006pk}, the following expressions:
\be
\label{LineIntegrals}
Q_s^2(\textbf{r}_0,\phi)=K\int_0^\infty d\xi\ \xi\ T^4(\xi)\ , \quad
\omega_m(\textbf{r}_0,\phi)=K\int_0^\infty d\xi\ \xi^2\ T^4(\xi)\ .
\ee
As done in standard weak-coupling calculations, the normalization will be fitted on
$R_{AA}$ data as a function of transverse hadron momentum $P_T$ for 10\% central 200 AGeV Au-Au collisions for one value of $P_T$. This value of $K$ is then used to predict the $P_T$ dependence, with a different hydrodynamical run the centrality dependence and also other observables such as back-to-back correlations.

Let us make a final comment. At weak coupling, gluons accumulate transverse momentum via a multiple scattering process characterized by the constant rate $dp^2_\perp/dt=\hat{q}\sim T^3,$ independent of the length of the medium (they receive random $p_\perp$ kicks). At strong coupling, the (faster) accumulation of transverse momentum can also be viewed as a multiple scattering process, but characterized by the constant rate $d|p_\perp|/dt=T^2$ (gluons receive $|p_\perp|$ kicks). Therefore, $dp^2_\perp/dt\sim T^2 |p_\perp|$ ($\to T^2 Q_s$ for the dominant fluctuations) is not a constant and should not be thought of a strong coupling result for the jet quenching parameter
$\hat{q}.$ For a finite-length plasma one has $dp^2_\perp/dt\propto T^4 L$ while for an infinite-length plasma the result is $dp^2_\perp/dt\propto T^3 \sqrt{E/M}$ (we recover the results of \cite{Gubser:2006nz,CasalderreySolana:2007qw,Giecold:2009cg}). Even though for infinite-extent matter the result looks similar to the weak coupling result, the physics is different. 

\section{Implementing the plasma geometry and expansion}

In order to compute observables from the quenching weight Eq.~(\ref{QuenchingWeight}), we have to evaluate this expression for any given path of a parton through the medium using the line integrals in Eqs.~(\ref{LineIntegrals}), average over the paths through the medium characteristic for a particular high$-P_T$ observable and finally fold the averaged energy loss probability with the QCD expressions for hard hadron production. In the following, the medium evolution is always specified in terms of a 3-d hydrodynamical model \cite{Hydro} which is well constrained by a large number of bulk observables. From this model, the local medium temperature $T$ at any spacetime point can be determined for Eqs.~(\ref{LineIntegrals}). For comparisons, we will also use Eqs.~(\ref{weakLI}). In this case $T$ will be substituted by $\epsilon^{1/4}$ where $\epsilon$ is the energy density to make the exact connection to previously published results where this scaling was assumed. This is known to make a difference in terms of fitting $K$ but it does not change the shape of the resulting suppression curves, see \cite{Bass:2008rv} for a detailed discussion.

The computation is for technical reasons different for single hadron suppression in terms of the nuclear modification factor $R_{AA}$ and for dihadron correlations in terms of the suppression factor $I_{AA}$. While the former quantity is obtained with standard multi-dimensional numerical integration techniques, the latter quantity is solved in a Monte Carlo (MC) simulation. A detailed description of the model is found in \cite{Renk:2006pk}, here we outline the essential steps.

From the quenching weight $P(\Delta E)$ given a single path in the medium as computed using
Eqs.~(\ref{LineIntegrals}), we define the averaged energy loss probability distribution
$\langle P(\Delta E)\rangle_{T_{AA}}$ as
\begin{equation}
\label{E-P_TAA}
\langle P(\Delta E)\rangle_{T_{AA}} \negthickspace = \negthickspace \frac{1}{2\pi} \int_0^{2\pi}  
\negthickspace \negthickspace \negthickspace d\phi 
\int_{-\infty}^{\infty} \negthickspace \negthickspace \negthickspace \negthickspace dx_0 
\int_{-\infty}^{\infty} \negthickspace \negthickspace \negthickspace \negthickspace dy_0 P(x_0,y_0)  
P(\Delta E).
\end{equation}
The averaging weight $P(x_0, y_0)$, i.e. the probability  for finding a hard vertex at the 
transverse position ${\bf r_0} = (x_0,y_0)$ and impact 
parameter ${\bf b},$ is given by the product of the nuclear profile functions:
\begin{equation}
\label{E-Profile}
P(x_0,y_0) = \frac{T_{A}({\bf r_0 + b/2}) T_A(\bf r_0 - b/2)}{T_{AA}({\bf b})},
\end{equation}
where the thickness function is given in terms of Woods-Saxon the nuclear density
$\rho_{A}({\bf r},z)$ as $T_{A}({\bf r})=\int dz \rho_{A}({\bf r},z)$.
The energy loss probability  $P(\Delta E)$ is derived in the limit of infinite parton energy. In order to account for the finite energy of the partons we truncate $\langle P(\Delta E) \rangle_{T_{AA}}$ 
at $\Delta E = E_{\rm jet}$ and add $\delta(\Delta-E_{\rm jet}) \int^\infty_{E_{\rm jet}} d\epsilon P(\epsilon)$ to the truncated distribution to ensure proper normalization. The physics meaning of this correction is that we consider all partons as absorbed whose energy loss is formally larger than their initial energy.
We calculate the momentum spectrum of hard partons in leading order perturbative QCD (LO pQCD) (explicit expressions are given in \cite{Renk:2006pk} and references therein). The medium-modified perturbative production of hadrons can then be computed from the expression
\begin{equation}
d\sigma_{med}^{AA\rightarrow h+X} \negthickspace \negthickspace = \sum_f d\sigma_{vac}^{AA \rightarrow f +X} \otimes \langle P(\Delta E)\rangle_{T_{AA}} \otimes
D_{f \rightarrow h}^{vac}(z, \mu_F^2)
\end{equation} 
with $D_{f \rightarrow h}^{vac}(z, \mu_F^2)$ the fragmentation function with momentum fraction $z$ at scale $\mu_F^2$ \cite{AKK}, and from this we compute the nuclear modification factor $R_{AA}$  which is the ratio of hard hadron production in A-A collisions normalized to the production in p-p collisions scaled by the number of binary collisions, i.e.
\begin{equation}
R_{AA}(P_T,y) = \frac{dN^h_{AA}/dP_Tdy }{T_{AA}({\bf b}) d\sigma^{pp}/dP_Tdy}.
\end{equation}

For the computation of in-medium back-to-back correlations with a hard triggered hadron, the trigger bias has to be included consistently into the averaging over geometry. For this purpose, we simulate a large number of back-to-back events in the medium. We start with the pQCD expression for the production of partons $k,l$ from two colliding objects $A,B$
\begin{equation}
\label{E-2Parton}
\frac{d\sigma^{AB\rightarrow kl +X}}{d p_T^2 dy_1 dy_2} \negthickspace = \sum_{ij} x_1 f_{i/A} 
(x_1, Q^2) x_2 f_{j/B} (x_2,Q^2) \frac{d\hat{\sigma}^{ij\rightarrow kl}}{d\hat{t}}
\end{equation}
where $f_{i/A}(x_1, Q^2)$ is the parton distribution function at fractional momentum $x_1$ and scale $Q^2$ which is different in the proton \cite{CTEQ1,CTEQ2} and in the nucleus \cite{NPDF,EKS98} and $\frac{d\hat{\sigma}^{ij\rightarrow kl}}{d\hat{t}}$ are perturbatively calculable cross-sections for the various hard partonic subprocesses. We MC sample this expression at midrapidity $y_1 = y_2 = 0$ to get a back-to-back parton pair with given parton type and momentum $p_T$. At the same time, the spatial position of the event is MC sampled from Eq.~(\ref{E-Profile}). We propagate each of the partons through the medium and compute $P(\Delta E)$ using Eqs.(\ref{LineIntegrals}), then determine probabilistically a value of $\Delta E$ from $P(\Delta E)$ and subtract it from the parton energy before hadronization. If $\Delta E > E$, i.e. if the parton energy loss is larger than its energy, we consider the parton absorbed by the medium, otherwise it emerges with a finite energy and we hadronize it subsequently.

In order to determine if there is a trigger hadron above a given threshold, given a parton $k$ with momentum $p_T$, we need to sample $A_1^{k\rightarrow h}(z_1, p_T)$, i.e. the probability distribution to find a hadron $h$ from the parton $k$ where $h$ is the most energetic hadron of the shower and carries the momentum $P_T = z_1 \cdot p_T$. We extract $A_1(z_1, p_T)$ from the shower evolution code HERWIG \cite{HERWIG}. The procedure is described in detail in \cite{dihadron-LHC}. Sampling $A_1(z_1, p_T)$ for any parton which emerged with sufficient energy from the medium provides the energy of the two most energetic hadrons on both sides of the event. The harder of these two defines the near side. The hadron opposite to it is then the leading away side hadron. In order to compute the correlation strength associated with subleading fragmentation of a parton emerging from the medium we evaluate $A_2(z_1, z_2, p_T)$ (also extracted from HERWIG), the conditional probability to find the second most energetic hadron at momentum fraction $z_2$ {\em given that the most energetic hadron was found with fraction $z_1$}. Given a trigger, we can then count the number of events in which correlated hadrons on near and away side fall into certain momentum windows and determine the per-trigger yield. The dihadron suppression factor $I_{AA}$ is then computed by dividing the per-trigger yield in A-A collisions by the per-trigger yield in p-p collisions.

\section{Results}

\begin{figure*}
\epsfig{file=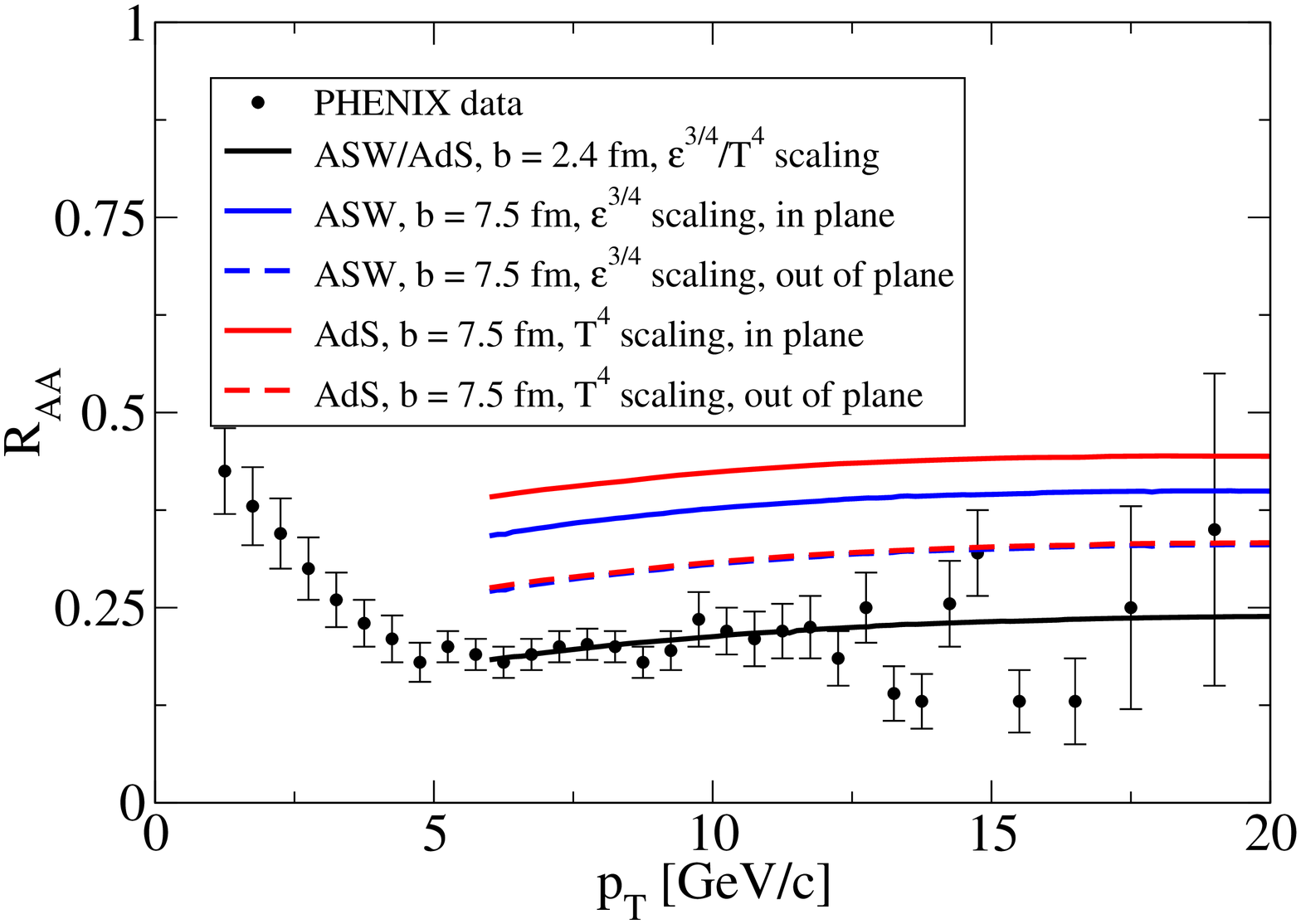, width=8cm} \epsfig{file=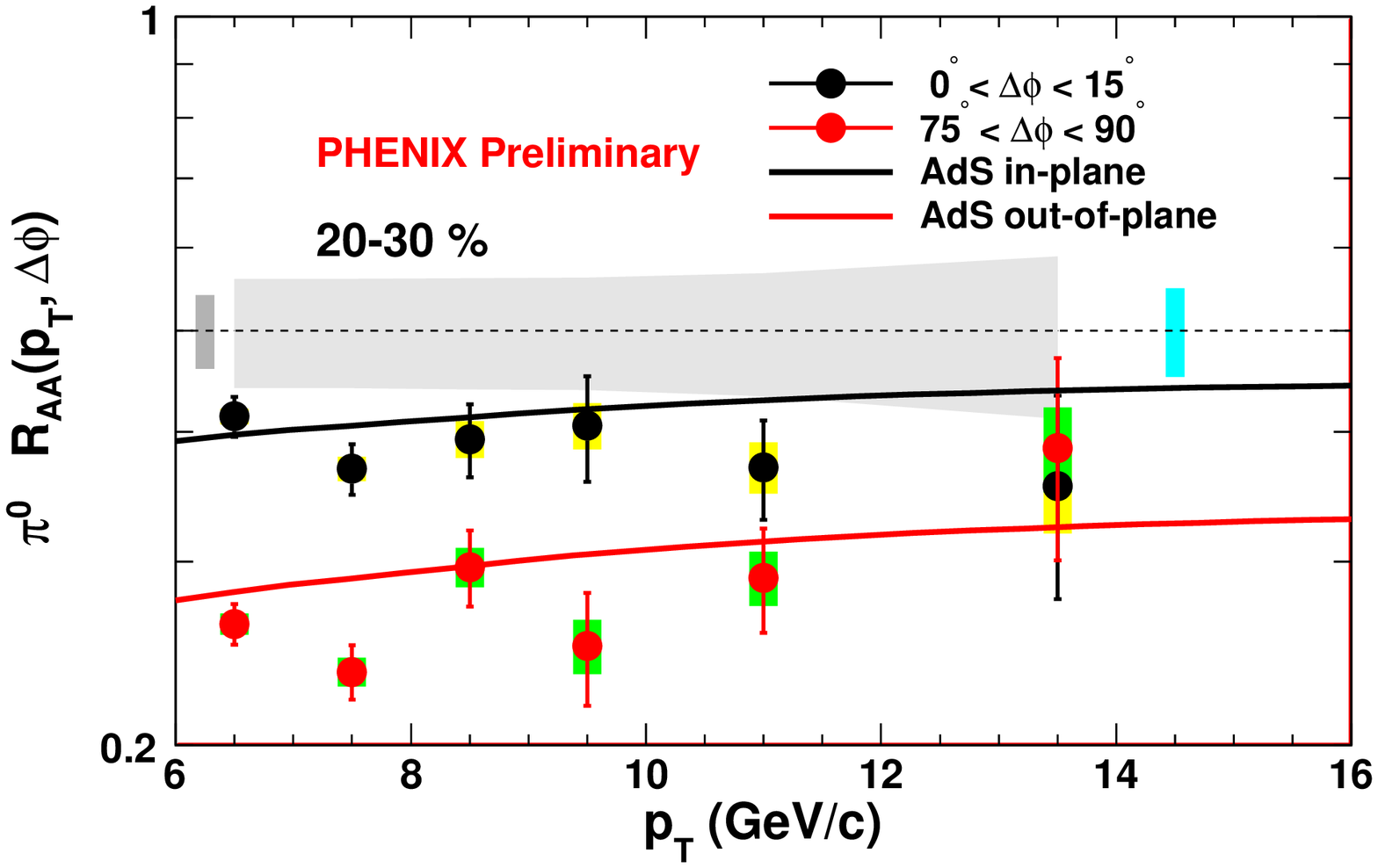, width=9cm}
\caption{\label{F-RAA}Left panel: Nuclear modification factor $R_{AA}$ for 200 AGeV Au+Au collisions as a function of $P_T,$ for central collisions (data and $b=2.4$ fm) and 25\% peripheral collisions ($b=7.5$ fm). Results for the weak coupling (labeled ASW) and strong coupling
(labeled AdS) scenarios for the medium are displayed (see text). The data are from \cite{PHENIX_R_AA}. Right panel: the same strong coupling predictions are compared with data for peripheral collisions and two reaction plane angles.
The data are from \cite{PHENIX_preliminary-Carla,PHENIX_preliminary-Rui}.}
\end{figure*}

We show the $P_T$ dependence of the nuclear modification factor $R_{AA}$ for 200 AGeV Au+Au collisions as compared with the PHENIX data for 10\% central collisions \cite{PHENIX_R_AA} in
Fig.~\ref{F-RAA} (left plot) for both the weak coupling and the strong coupling assumption for the medium. In both cases, with a single fit parameter, a good (and virtually identical) description of the data can be made. However, in the extrapolation to more peripheral collisions (impact parameter $b=7.5$ fm), differences between the two scenarios become apparent, dependent on the angular orientation of the high$-P_T$ hadron with respect to the reaction plane. In particular, the spread between in-plane and out-of-plane emissions is increased in the strong coupling case, which is in agreement with preliminary data \cite{PHENIX_preliminary-Carla,PHENIX_preliminary-Rui} as is shown in Fig.~\ref{F-RAA} (right plot).

This is consistent with the different pathlength dependence of energy loss which is for a constant medium $L^2$ in the weak coupling case but $L^3$ in the strong coupling case. Consequently, the average pathlength difference between in-plane and out-of-plane emission is weighted with a different power also in the case of the expanding hydrodynamical medium, which translates into a stronger effect in the strong coupling case. Because both calculations are adjusted on central collisions, they predict a similar suppression for out-of-plane emissions (the $R_{AA}$'s are similar to each other but both bigger than for central collisions where the temperature is higher), as these involve the same average pathlength. Consequently, the stronger $L$ dependence causes the strong coupling scenario to predict less suppression than the weak-coupling scenario for in-plane emissions, where the pathlength is on average shorter.

\begin{figure*}
\epsfig{file=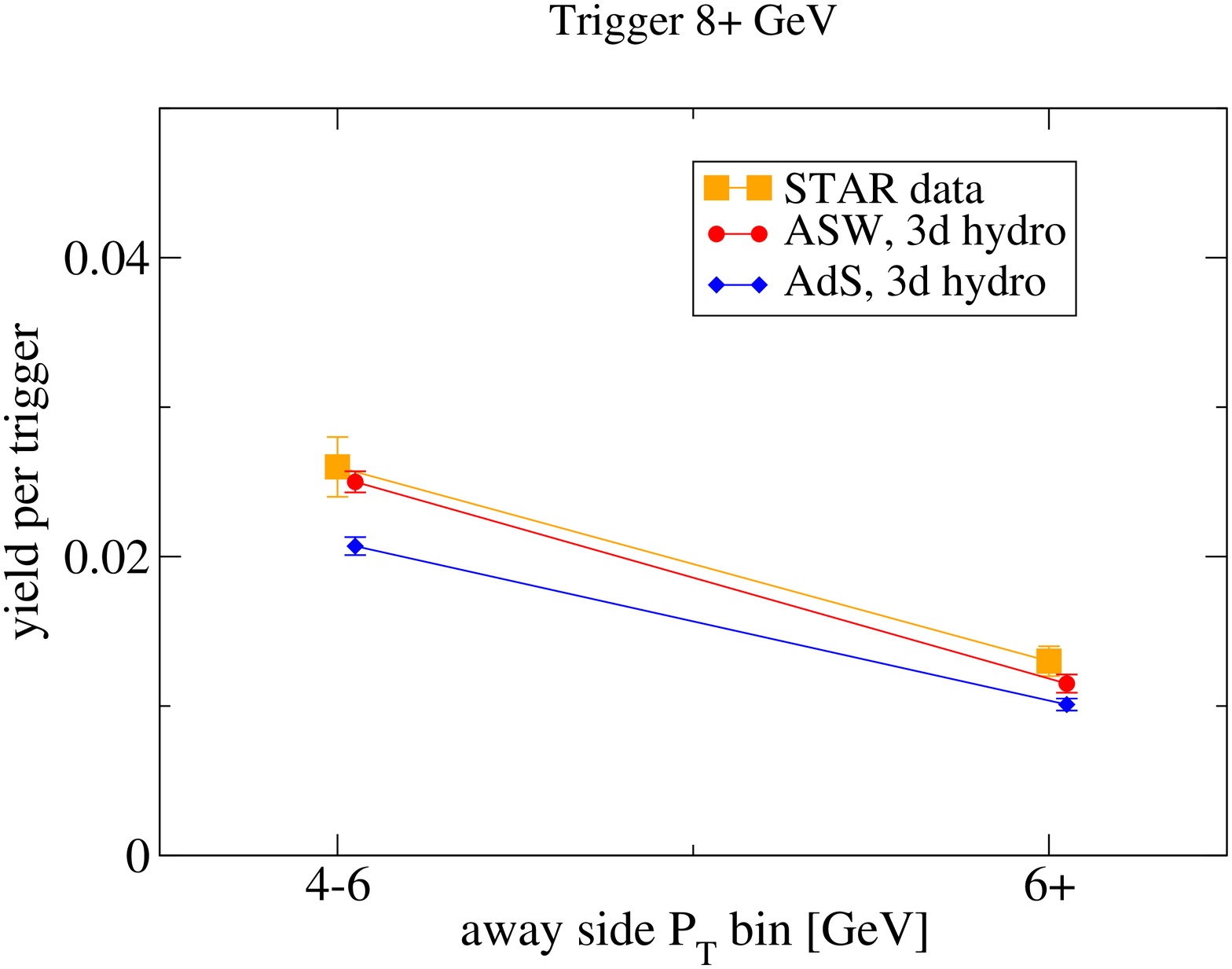, width=8cm} \epsfig{file=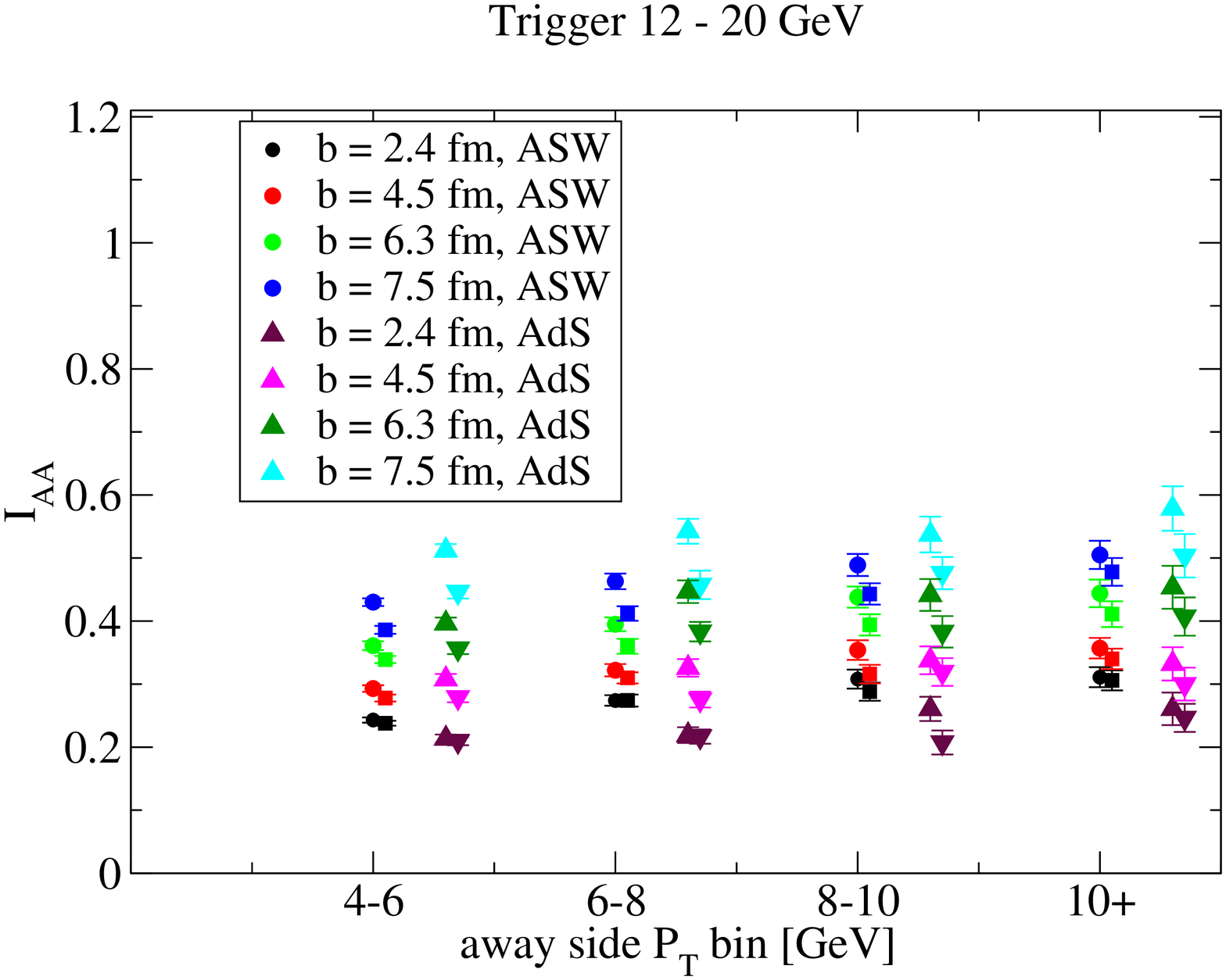, width=8cm}
\caption{\label{F-B2B}Left panel: Per trigger yield on the away side with respect to a 8+ GeV hadron trigger in back-to-back correlations in central 200 AGeV Au+Au collisions for two different momentum bins. Shown are STAR data \cite{STAR_Dijets}, the calculated result for a weak coupling and for a strong coupling treatment of the medium. Right panel: Nuclear dihadron away side suppression factor $I_{AA}$ for a 12-20 GeV trigger in 200 AGeV Au+Au collisions for different impact parameter and in-plane vs. out-of-plane emission as a function of away-side momentum bins. Shown are weak coupling (ASW) and strong coupling (AdS) treatment of the medium. The points have been shifted along the $x$-direction within each momentum bin for clarity of presentation.}
\end{figure*}

Based on the results of \cite{Elastic}, back-to-back correlations where the difference in average pathlength is maximized due to the surface bias of the trigger hadron would be an appropriate observable to show the differences between the weak and strong coupling scenario for the medium more clearly. The away side per-trigger yield for an 8+ GeV hadron trigger as measured in 200 AGeV central Au-Au collisions by the STAR collaboration \cite{STAR_Dijets} is shown in the left panel of Fig.~\ref{F-B2B} and compared with both the weak coupling and the strong coupling treatment of the medium. As expected, the different weighting of the pathlength difference in the two approaches translates into a stronger suppression for the $L^3$ pathlength dependence of the strong coupling approach, the net effect however is mild, especially in the region of associate hadrons at 6+ GeV momentum where fragmentation can safely be assumed to dominate hadron production, and also where our approach is best applicable.

In the right panel of Fig.\ref{F-B2B}, we present a more differential observable: the away side dihadron suppression factor $I_{AA}$ for various collision centralities and away side momentum bins, as calculated in \cite{Dihadron_ang} within the weak coupling approach to medium properties, and compared with the strong coupling approach of this letter. The presentation within a bin is arranged in two groups, the weak coupling results first, the strong-coupling results next. In each group, the color coding refers to the centrality of the hydro run (there is more suppression for more central collisions), and for each centrality there are two points (with different symbol shape) - the in-plane result is always the upper point, showing less suppression, the out of plane result is the lower point. The in-plane and out-of-plane results are closer to each other for more central collisions. Finally, one observes that the strong coupling approach predicts more suppression than the weak coupling one for central collisions (consistent with the per-trigger yield result presented above) but less suppression for peripheral collisions, and a larger spread between in-plane and out-of-plane emissions.

For completeness, the values of the parameter adjusted to describe $R_{AA}$ for central collisions are $K\simeq 27$ in the weak-coupling case (Eqs.~(\ref{weakLI})) and $K\simeq 6$ in the strong-coupling case (Eqs.~(\ref{LineIntegrals})). The weak-coupling value is a factor 3.5 bigger than expected for an ideal gas \cite{Baier:2002tc}; we are not aware of a formula with which we could compare the strong-coupling result, but the fact that $K$ is of order one is satisfactory.

\section{Conclusions}

In this letter, we propose a new description of the quenching of high$-p_T$ particles moving in a strongly-interacting QGP. In our model, energetic partons lose energy through the emission of long-lived gluons (with coherence time $t_c\sim L,$ the plasma extent), as dictated by perturbative calculations, while the amount of energy that these gluons carry away in the medium is enhanced by the strongly-coupled nature of the plasma, compared to weak-coupling calculations. We do not perform a direct AdS/CFT calculation, we rather follow the idea of Ref.~\cite{LRW}, that whether the plasma is weakly-coupled or not, gluon radiation is the dominant mechanism by which high$-p_T$ particles lose energy. However, the assumption in \cite{LRW} to work with a constant $\hat{q}$ computed at strong-coupling does not allow to implement the path-length dependence of the energy loss, which is important for phenomenology. Our prescription does.

With respect to a weakly-coupled plasma, the path length dependence of the energy loss of energetic partons in a strongly-coupled plasma is stronger ($\omega_m\!=\!T^4L^3$ vs.
$\omega_m\!=\!\hat{q}L^2$). In order to quantify the effect that this has on physical observables like $R_{AA}$ and $I_{AA}$ at high $P_T$, we have proposed to substitute Eqs.(\ref{weakLI}) in a successful phenomenological analysis at weak coupling, by their strong-coupling expressions Eqs.(\ref{LineIntegrals}). In both cases we have assumed $\alpha_s(p_T^2)\ll1,$ the high-parton energy limit, and used the BDMPS-Z radiative energy loss formalism. Only for low$-p_T$ observables can one hope to apply strong-coupling techniques to the entire process, and interesting results for dihadron correlations have also been obtained in this case
\cite{Noronha:2008un,Betz:2008wy}.

Our two calculations are fixed by one-parameter fits to $R_{AA}$ for central Au+Au collisions, resulting in good and comparable descriptions of the data. Differences between the two cases are then seen when considering non-central collisions: the spread between the in-plane and out-of-plane $R_{AA}$'s is increased in the strong coupling case, which is in good agreement with the data. The suppression of $I_{AA}$ is also increased for central collisions, as well as the spread between the in-plane and out-of-plane values for non-central collisions. Our hybrid model allows to make quantitative predictions which could be tested with future measurements. One concern is that the theoretical uncertainties could be large enough to blur the potential resolving power of the proposed measurements, for which the quantitative differences between the two approaches are small. Moreover, since a major source of uncertainty in jet quenching calculations (and more globally in heavy-ion collisions) is the shape of the initial hydro profile (CGC vs. Glauber), quantifying the theoretical uncertainty is not straightforward.

In the future, it would also be interesting to investigate how our substitution impacts the $R_{AA}$ of heavy hadrons (or rather non-photonic electrons), which is difficult to accommodate in the pQCD framework, and one could also compare with results obtained fully from strong-coupling dynamics \cite{Horowitz:2007su}.

\begin{acknowledgments}

We are grateful to Barbara Jacak and the PHENIX collaboration for allowing us to show their preliminary data. In addition we thank Carla Vale and Rui Wei for making Fig.1 (right).
C.M. would like to thank Al Mueller for discussions and reading the manuscript. 
C.M. is supported by the European Commission under the FP6 program, contract No. MOIF-CT-2006-039860, T.R. acknowledges support from the Academy of Finland, Project 115262. 

\end{acknowledgments}


\begin{thebibliography}{99}

\bibitem{brahms}
  I.~Arsene {\it et al.}  [BRAHMS Collaboration],
  Nucl.\ Phys.\  A {\bf 757}, 1 (2005).

\bibitem{phobos}
  B.~B.~Back {\it et al.}  [PHOBOS Collaboration],
  Nucl.\ Phys.\  A {\bf 757}, 28 (2005).

\bibitem{star}
  J.~Adams {\it et al.}  [STAR Collaboration],
  Nucl.\ Phys.\  A {\bf 757}, 102 (2005).

\bibitem{phenix}
  K.~Adcox {\it et al.}  [PHENIX Collaboration],
  Nucl.\ Phys.\  A {\bf 757}, 184 (2005).

\bibitem{Dusling:2007gi}
  K.~Dusling and D.~Teaney,
  Phys.\ Rev.\  C {\bf 77}, 034905 (2008).

\bibitem{Song:2007ux}
  H.~Song and U.~W.~Heinz,
  Phys.\ Rev.\  C {\bf 77}, 064901 (2008).

\bibitem{Luzum:2008cw}
  M.~Luzum and P.~Romatschke,
  Phys.\ Rev.\  C {\bf 78}, 034915 (2008)
  [Erratum-ibid.\  C {\bf 79}, 039903 (2009)].

\bibitem{Baier:1998kq}
  R.~Baier, Y.~L.~Dokshitzer, A.~H.~Mueller and D.~Schiff,
  Nucl.\ Phys.\  B {\bf 531}, 403 (1998).

\bibitem{Wiedemann:2000za}
  U.~A.~Wiedemann,
  Nucl.\ Phys.\  B {\bf 588}, 203 (2000).

\bibitem{Gyulassy:2000er}
  M.~Gyulassy, P.~Levai and I.~Vitev,
  Nucl.\ Phys.\  B {\bf 594}, 371 (2001).

\bibitem{Eskola:2004cr}
  K.~J.~Eskola, H.~Honkanen, C.~A.~Salgado and U.~A.~Wiedemann,
  Nucl.\ Phys.\  A {\bf 747}, 511 (2005).

\bibitem{Bass:2008rv}
  S.~A.~Bass, C.~Gale, A.~Majumder, C.~Nonaka, G.~Y.~Qin, T.~Renk and J.~Ruppert,
  Phys.\ Rev.\  C {\bf 79}, 024901 (2009).

\bibitem{Maldacena:1997re}
  J.~M.~Maldacena,
  Adv.\ Theor.\ Math.\ Phys.\  {\bf 2}, 231 (1998)
  [Int.\ J.\ Theor.\ Phys.\  {\bf 38}, 1113 (1999)].
  
\bibitem{Witten:1998qj}
  E.~Witten,
  Adv.\ Theor.\ Math.\ Phys.\  {\bf 2}, 253 (1998).
  
\bibitem{Gubser:1998bc}
  S.~S.~Gubser, I.~R.~Klebanov and A.~M.~Polyakov,
  Phys.\ Lett.\  B {\bf 428}, 105 (1998).

\bibitem{Policastro:2001yc}
  G.~Policastro, D.~T.~Son and A.~O.~Starinets,
  Phys.\ Rev.\ Lett.\  {\bf 87}, 081601 (2001).

\bibitem{Kovtun:2004de}
  P.~Kovtun, D.~T.~Son and A.~O.~Starinets,
  Phys.\ Rev.\ Lett.\  {\bf 94}, 111601 (2005).

\bibitem{Kovner:2003zj}
  A.~Kovner and U.~A.~Wiedemann,
  arXiv:hep-ph/0304151.

\bibitem{LRW}
  H.~Liu, K.~Rajagopal and U.~A.~Wiedemann,
  Phys.\ Rev.\ Lett.\  {\bf 97}, 182301 (2006);
  JHEP {\bf 0703}, 066 (2007).

\bibitem{BDMPS}
  R.~Baier, Y.~L.~Dokshitzer, A.~H.~Mueller, S.~Peigne and D.~Schiff,
  Nucl.\ Phys.\  B {\bf 483}, 291 (1997);
  {\it ibid.}\ {\bf 484}, 265 (1997).

\bibitem{Zakharov}
  B.~G.~Zakharov,
  JETP Lett.\  {\bf 63}, 952 (1996);
  {\it ibid.}\ {\bf 65}, 615 (1997).

\bibitem{Mueller:2008zt}
  A.~H.~Mueller,
  Phys.\ Lett.\  B {\bf 668}, 11 (2008).

\bibitem{Dominguez:2008vd}
  F.~Dominguez, C.~Marquet, A.~H.~Mueller, B.~Wu and B.~W.~Xiao,
  Nucl.\ Phys.\  A {\bf 811}, 197 (2008).

\bibitem{Marquet:2008kr}
  C.~Marquet,
  Eur.\ Phys.\ J.\  C {\bf 62}, 15 (2009).

\bibitem{Renk:2006pk}
  T.~Renk and K.~Eskola,
  Phys.\ Rev.\  C {\bf 75}, 054910 (2007).

\bibitem{Salgado:2003gb}
  C.~A.~Salgado and U.~A.~Wiedemann,
  Phys.\ Rev.\  D {\bf 68}, 014008 (2003).

\bibitem{Wiedemann:2000tf}
  U.~A.~Wiedemann,
  Nucl.\ Phys.\  A {\bf 690}, 731 (2001).

\bibitem{Hatta:2007cs}
  Y.~Hatta, E.~Iancu and A.~H.~Mueller,
  JHEP {\bf 0801}, 063 (2008).

\bibitem{Mueller:2008bt}
  A.~H.~Mueller, A.~I.~Shoshi and B.~W.~Xiao,
  Nucl.\ Phys.\  A {\bf 822}, 20 (2009).

\bibitem{Levin:2009vj}
  E.~Levin, J.~Miller, B.~Z.~Kopeliovich and I.~Schmidt,
  JHEP {\bf 0902}, 048 (2009).

\bibitem{Albacete:2008ze}
  J.~L.~Albacete, Y.~V.~Kovchegov and A.~Taliotis,
  JHEP {\bf 0807}, 074 (2008).

\bibitem{CasalderreySolana:2007sw}
  J.~Casalderrey-Solana and X.~N.~Wang,
  Phys.\ Rev.\  C {\bf 77}, 024902 (2008).

\bibitem{Gubser:2008as}
  S.~S.~Gubser, D.~R.~Gulotta, S.~S.~Pufu and F.~D.~Rocha,
  JHEP {\bf 0810}, 052 (2008).

\bibitem{Chesler:2008uy}
  P.~M.~Chesler, K.~Jensen, A.~Karch and L.~G.~Yaffe,
  Phys.\ Rev.\  D {\bf 79}, 125015 (2009).

\bibitem{Dokshitzer:2001zm}
  Y.~L.~Dokshitzer and D.~E.~Kharzeev,
  Phys.\ Lett.\  B {\bf 519}, 199 (2001).

\bibitem{Herzog:2006gh}
  C.~P.~Herzog, A.~Karch, P.~Kovtun, C.~Kozcaz and L.~G.~Yaffe,
  JHEP \textbf{0607}, 013 (2006).  

\bibitem{Gubser:2006bz}
  S.~S.~Gubser,
  Phys.\ Rev.\  D {\bf 74}, 126005 (2006).

\bibitem{Chernicoff:2008sa}
  M.~Chernicoff and A.~Guijosa,
  JHEP {\bf 0806}, 005 (2008).

\bibitem{Beuf:2008ep}
  G.~Beuf, C.~Marquet and B.~W.~Xiao,
  Phys.\ Rev.\  D {\bf 80}, 085001 (2009).

\bibitem{Gubser:2006nz}
  S.~S.~Gubser,
  Nucl.\ Phys.\  B {\bf 790}, 175 (2008).

\bibitem{CasalderreySolana:2007qw}
  J.~Casalderrey-Solana and D.~Teaney,
  JHEP {\bf 0704}, 039 (2007);
  Phys.\ Rev.\  D {\bf 74}, 085012 (2006).

\bibitem{Giecold:2009cg}
  G.~C.~Giecold, E.~Iancu and A.~H.~Mueller,
  JHEP {\bf 0907}, 033 (2009).

\bibitem{Hydro}
  C.~Nonaka and S.~A.~Bass,
  Phys.\ Rev.\ C {\bf 75}, 014902 (2007).

\bibitem{AKK}
  S.~Albino, B.~A.~Kniehl and G.~Kramer,
  Nucl.\ Phys.\ B {\bf 725}, 181 (2005).

\bibitem{CTEQ1}
  J.~Pumplin, D.~R.~Stump, J.~Huston, H.~L.~Lai, P.~M.~Nadolsky and W.~K.~Tung,
  JHEP {\bf 0207}, 012 (2002).
 
\bibitem{CTEQ2}
  D.~Stump, J.~Huston, J.~Pumplin, W.~K.~Tung, H.~L.~Lai, S.~Kuhlmann and J.~F.~Owens,
  JHEP {\bf 0310}, 046 (2003).

 \bibitem{NPDF}
  M.~Hirai, S.~Kumano and T.~H.~Nagai,
  Phys.\ Rev.\  C {\bf 70}, 044905 (2004).

\bibitem{EKS98}
  K.~J.~Eskola, V.~J.~Kolhinen and C.~A.~Salgado,
  Eur.\ Phys.\ J.\  C {\bf 9}, 61 (1999).

\bibitem{HERWIG}
  G.~Corcella {\it et al.},
  arXiv:hep-ph/0210213.

\bibitem{dihadron-LHC}
  T.~Renk and K.~J.~Eskola,
  Phys.\ Rev.\  C {\bf 77}, 044905 (2008).

\bibitem{PHENIX_R_AA}
  M.~Shimomura  [PHENIX Collaboration],
  Nucl.\ Phys.\  A {\bf 774}, 457 (2006).

\bibitem{PHENIX_preliminary-Carla}
  C.~M.~Vale, for the PHENIX Collaboration,
  arXiv:0907.4729 [nucl-ex].

\bibitem{PHENIX_preliminary-Rui}
  R.~Wei, for the PHENIX Collaboration,
  arXiv:0907.0024 [nucl-ex].

\bibitem{Baier:2002tc}
  R.~Baier,
  Nucl.\ Phys.\  A {\bf 715}, 209 (2003).

\bibitem{Elastic}
  T.~Renk,
  Phys.\ Rev.\  C {\bf 76}, 064905 (2007).

\bibitem{STAR_Dijets}
 J.~Adams {\it et al.}  [STAR Collaboration],
  Phys.\ Rev.\ Lett.\  {\bf 97}, 162301 (2006).

\bibitem{Dihadron_ang}
  T.~Renk,
  Phys.\ Rev.\  C {\bf 78}, 034904 (2008).

\bibitem{Noronha:2008un}
  J.~Noronha, M.~Gyulassy and G.~Torrieri,
  Phys.\ Rev.\ Lett.\  {\bf 102}, 102301 (2009).

\bibitem{Betz:2008wy}
  B.~Betz, M.~Gyulassy, J.~Noronha and G.~Torrieri,
  Phys.\ Lett.\  B {\bf 675}, 340 (2009).

\bibitem{Horowitz:2007su}
  W.~A.~Horowitz and M.~Gyulassy,
  Phys.\ Lett.\  B {\bf 666}, 320 (2008).

\end{thebibliography}
\end{document}